\documentclass{article}
\usepackage{spconf,amsmath,graphicx,hyperref}

\usepackage{amsfonts}
\usepackage{amssymb}
\usepackage{subcaption}
\usepackage{siunitx}
\usepackage{booktabs}
\usepackage{tikz}
\usepackage{xcolor}
\usepackage{adjustbox}
\usetikzlibrary{arrows.meta, positioning, calc, shapes}
\usepackage{multirow}

\usepackage{comment}
\usepackage{algorithm}
\usepackage{algpseudocode}

\title{Improving Test-Time Performance of RVQ-based Neural Codecs}
\name{Hyeongju Kim$^{1}$ \qquad Junhyeok Lee$^{2}$ \qquad Jacob Morton$^{1}$ \qquad Juheon Lee$^{1}$ \qquad Jinhyeok Yang$^{1}$\thanks{Corresponding author: \texttt{hyeongju@supertone.ai}}}
  \address{$^{1}$Supertone Inc., Seoul, Republic of Korea \\
      $^{2}$Center for Language and Speech Processing, Johns Hopkins University, Baltimore, MD, USA}
\begin{document}
\ninept
\maketitle
\begin{abstract}
The residual vector quantization (RVQ) technique plays a central role in recent advances in neural audio codecs. These models effectively synthesize high-fidelity audio from a limited number of codes due to the hierarchical structure among quantization levels. In this paper, we propose an encoding algorithm to further enhance the synthesis quality of RVQ-based neural codecs at test-time. Firstly, we point out the suboptimal nature of quantized vectors generated by conventional methods. We demonstrate that quantization error can be mitigated by selecting a different set of codes. Subsequently, we present our encoding algorithm, designed to identify a set of discrete codes that achieve a lower quantization error. We then apply the proposed method to pre-trained models and evaluate its efficacy using diverse metrics. Our experimental findings validate that our method not only reduces quantization errors, but also improves synthesis quality.

\end{abstract}
\section{Introduction}
Neural codec models have exhibited impressive data compression performance across various domains, including audio~\cite{zeghidour2021soundstream, defossez2023high, kumar2024high}, image~\cite{yang2023computationally, wang2022neural, yang2020improving}, and video~\cite{Khani_2021_ICCV, habibian2019video, fathima2023neural}.
These models effectively identify and discard data features that are insignificant to human perception, enabling high-fidelity reconstruction while minimizing storage and transmission costs. 
One common approach for this purpose involves employing an encode-decoder framework with discrete latent variables~\cite{van2017neural}. 
An encoder transforms source data into low-dimensional discrete representations, known as codes, through a quantization process and a decoder reconstructs the original data from these compressed representations.

Particularly in the field of neural audio codec, a residual vector quantization~(RVQ) technique emerges as a cornerstone in recent advancements.
The RVQ technique, first introduced by SoundStream~\cite{zeghidour2021soundstream}, defines a discrete latent variable as a sum of vectors from a finite set of multiple codebooks. 
The quantization is conducted in a hierarchical manner, where each $l$-th quantization step approximates the difference between the output of the previous $(l-1)$-th quantization step and the original continuous data. 
This scheme induces low-level quantizers to prioritize modeling coarse features of input audio, while high-level quantizers focus on capturing fine features.
The hierarchical structure of RVQ-based neural codecs facilitates efficient training, enabling the reconstruction of source audio with high-fidelity. 
Furthermore, RVQ-based neural codecs provide the capability to control bit-rate during inference by allowing the selection of the number of quantization steps. 
These advantages lead the RVQ technique to be utilized in subsequent research across various audio applications such as text-to-speech~\cite{kharitonov2023speak, shen2023naturalspeech}, text-to-audio~\cite{kreuk2022audiogen, ziv2024masked}, and text-to-music~\cite{copet2024simple, lam2024efficient}, as well as neural audio codecs.
%subsequent audio application research such as text-to-speech, text-to-audio, and text-to-music, as well as neural audio codecs.
% and offer some degree of control over bit-rate during inference.
% by setting a hierarchical structure among codebooks.

In this paper, we seek to further improve the synthesis quality of RVQ-based neural codecs during test time. To this end, we first highlight the suboptimal property of conventional RVQ encoding algorithms. As each $l$-th quantization solely relies on the information from the previous $(l-1)$-th quantization output, the resulting sequence of codes is obtained in a greedy manner and may not represent an optimal solution across all potential quantization composites. Consequently, the quantization error could be further minimized by choosing an alternative sequence of codes. 
% We provide a straightforward example that illustrate this scenario for better understanding. 
Next, we propose an encoding algorithm to obtain a code sequence that has a lower quantization error. 
The proposed method utilizes a beam-search algorithm to approximate the optimal solution, effectively mitigating the substantial computational burden associated with exhaustively searching all possible composites. 
In the proposed algorithm, conventional RVQ encoding methods can be seen as a specific case that arises when the beam size is set to 1.
Also, the proposed encoding algorithm can generally be applied to existing RVQ-based models without requiring re-training.
In our experiments, the proposed algorithm is applied to two widely used neural audio codecs: EnCodec~\cite{defossez2023high} and HiFi-Codec~\cite{yang2023hifi}. 
We evaluate the efficacy of the proposed algorithm using diverse metrics such as quantization error, spectral distance, NISQA score~\cite{mittag2021nisqa}, PESQ~\cite{rix2001perceptual}, STOI~\cite{taal2010short}, SI-SNR. The experimental results strongly support the effectiveness of our algorithm, surpassing the conventional RVQ encoding methods.

\section{Related Work}
The field of audio codec research has witnessed significant advancements, particularly with the emergence of neural network-based approaches. SoundStream~\cite{zeghidour2021soundstream} offers efficient audio compression across various types of audio content at low bit-rates, maintaining high quality and supporting real-time processing. Notably, it introduces the innovative RVQ technique and employs quantizer dropout during model training to support variable bit-rates. Similarly, EnCodec~\cite{defossez2023high} utilizes the RVQ technique in an auto-encoder structure and is trained with multi-scale mel-spectrogram losses and adversarial losses. Also, a novel loss balancer mechanism adjusts the coefficients of each loss during training, effectively stabilizing the training process and enhancing performance. Kumar et al.~\cite{kumar2024high} further enhances the performance of neural audio codecs through the incorporation of periodic inductive biases, code lookup in a low-dimensional space, L2-normalized codes, and a quantizer dropout rate. HiFi-Codec proposes a group-residual vector quantization (GRVQ) technique requiring only four codebooks for high-quality reconstruction. The GRVQ technique splits a continuous variable into several groups, applies RVQ to each group, and merges the resulting quantized vectors into a single output.

\section{Residual Vector Quantization}
Vector quantization~(VQ) is a technique that discretizes a continuous input vector by identifying the code in a predefined codebook that best matches the input vector~\cite{van2017neural}. A common approach of utilizing multiple codebooks is to partition the input vector into groups and apply the VQ process to each group in parallel. In contrast, RVQ conducts quantization sequentially, depending on the quantization step. 
% To be specific, the $l$-th RVQ layer quantizes the residual computed from the original data and the $l-1$ quantization output. When $l=1$, RVQ quantizes the original data similarly to conventional VQ.
Specifically, each layer of RVQ quantizes the residual computed from the original data and the output of the previous quantization layer. When the layer index is 1, RVQ quantizes the original data in the same manner as conventional VQ. The final output is computed by summing all the quantized residuals obtained from each quantization step. 
Thanks to the hierarchical structure of RVQ, the quantized vectors progressively capture finer features as the quantization levels deepen, enabling diverse bit-rate encoding with a single neural codec.
\begin{figure}
    \centering
\includegraphics[width=0.97\linewidth]{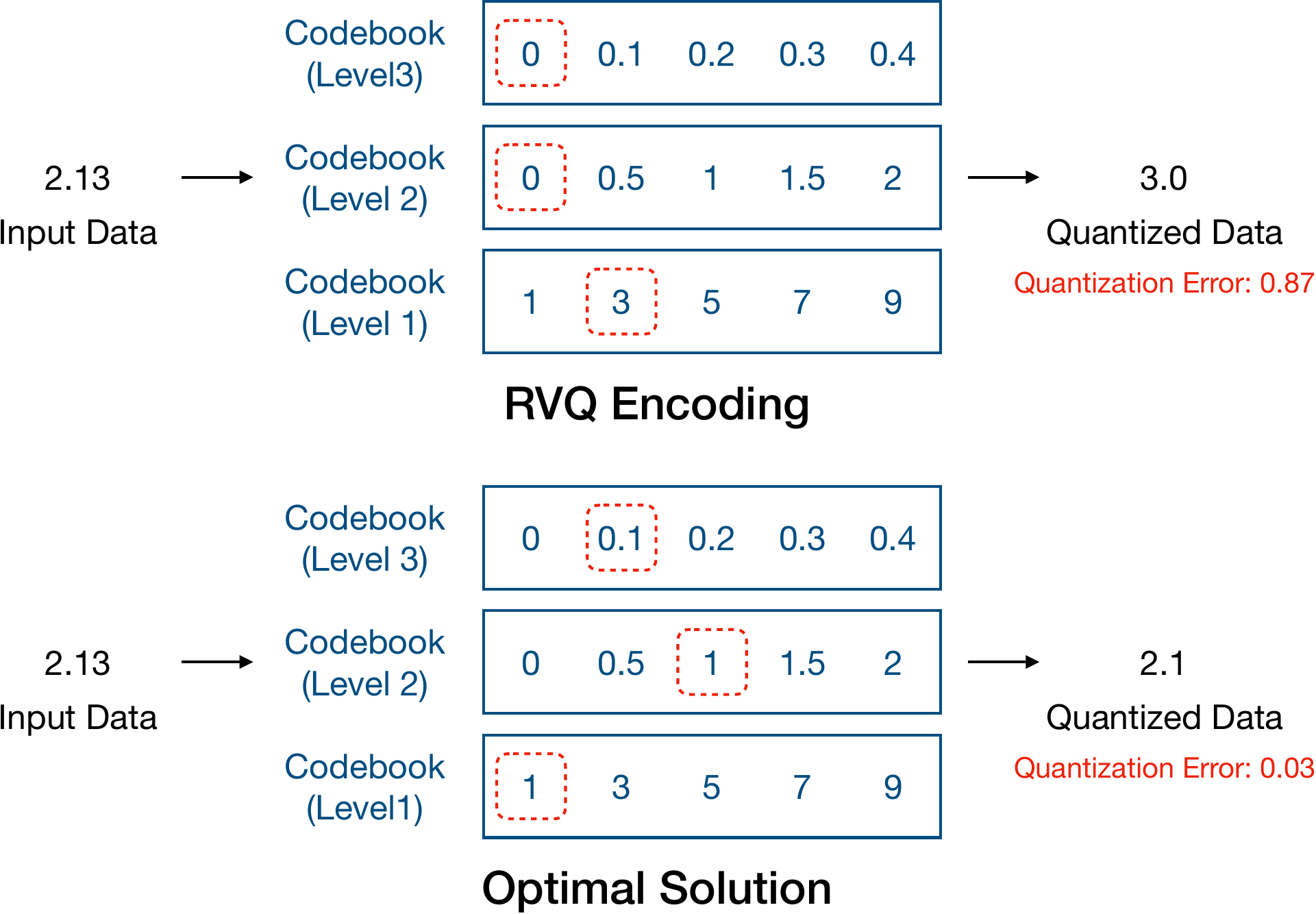} 
    \caption{Limitation of conventional RVQ encoding.}
\label{fig:fig1}
\end{figure}

However, conventional RVQ encoding algorithms have limitations in terms of quantization error. As each quantization step depends solely on the preceding output, the resulting sequence of codes may fail to represent an optimal solution across all potential quantization combinations. To illustrate this point, we provide an intuitive example in Fig.~\ref{fig:fig1}. Let's consider quantizing an input data value of 2.13 using conventional RVQ encoding with three codebooks. In the initial quantization level, the value 3 is chosen as it is the closest to 2.13 among all available options. In subsequent levels, the residual value of 0.87 undergoes further quantization steps by searching the best matching code in each codebook. The resulting RVQ encoding yields 3.0, with a quantization error of 0.87. However, the optimal code sequence that minimizes a quantization error is [1, 1, 0.1], resulting in an encoding value of 2.1 with an error of 0.03. This example suggests that a quantization error of RVQ encoding can be reduced by choosing a suitable code sequence, thereby improving the codec performance. 

The challenge lies in the fact that exploring all potential code combinations is impractical. Assuming the number of codebooks is denoted as $L$ and the size of each codebook as $S$, the computational cost escalates exponentially, following a complexity of $\mathcal{O}(S^{L})$. To circumvent this issue, we propose a new encoding algorithm in the next section.

\section{Proposed Method}
\begin{algorithm}[t]
\caption{Proposed Encoding Algorithm for RVQ}\label{alg:encoding}
\begin{algorithmic}[1]
\Require Input data $x$, beam size $B$, codebooks $C_i$ for $i=1..L$, search parameter $k$
\For{$i=1$ \textbf{to} $L$}
\If{$i = 1$}
\State Initialize an empty list $beam_1$
\State Find the top $B$ closest values to $x$ in $C_1$ and add them to $beam_1$
\Else
\State Initialize empty lists, $beam_i$ and $E_i$
\For{$j=1$ \textbf{to} $B$}
\State Compute the residual $r_j^i = x - beam_{i-1}[j]$
\State Find the top $k$ closest values to $r_j^i$ in $C_i$ and store them in $C_{\text{top}}$
\For{\textbf{each} $c$ in $C_{\text{top}}$}
\State Compute the new node $n = beam_{i-1}[j] + c$
\State Compute the quantization error $e = | x - n |^2$
\State Add $n$ to $beam_i$
\State Add $e$ to $E_i$
\EndFor
\EndFor
\State Sort nodes in $beam_i$ based on $E_i$
\State Prune $beam_i$ to keep only the top $B$ nodes
\EndIf
\EndFor
% \State $\hat{x} \gets $ The best node with minimum distortion from $beam_L$
\State $\hat{x} \gets$ The first node in $beam_L$
% \State $\hat{x} \gets beam_L[1]$
\State \Return{$\hat{x}$}
\end{algorithmic}
\end{algorithm}

\begin{table*}[]
\caption{Performance comparison of reconstructed speech across various beam sizes on diverse evaluation metrics.}
\label{tab:speech}
\centering
\begin{tabular}{@{\hspace{3mm}}clccccc@{\hspace{3mm}}}
\toprule
\multicolumn{1}{l}{}        &          & Mel Distance ↓         & SI-SNR ↑               & PESQ ↑                 & STOI ↑                 & NISQA ↑                \\ \midrule
\multirow{4}{*}{EnCodec}    & $B$ = 1  & 1.893 ± 0.029          & 4.239 ± 0.137          & 2.726 ± 0.010          & 0.942 ± 0.002          & 3.491 ± 0.022          \\ \cmidrule(l){2-7} 
                            & $B$ = 4  & 1.864 ± 0.029          & 4.521 ± 0.137          & 2.812 ± 0.010          & 0.946 ± 0.001          & 3.561 ± 0.022          \\
                            & $B$ = 8  & 1.852 ± 0.029          & 4.599 ± 0.137          & 2.836 ± 0.010          & 0.947 ± 0.001          & 3.570 ± 0.022          \\
                            & $B$ = 16 & \textbf{1.842 ± 0.029} & \textbf{4.649 ± 0.138} & \textbf{2.850 ± 0.010} & \textbf{0.948 ± 0.001} & \textbf{3.588 ± 0.022} \\ \midrule
\multirow{4}{*}{HiFi-Codec} & $B$ = 1  & 0.900 ± 0.006          & 3.711 ± 0.129          & 3.013 ± 0.010          & 0.948 ± 0.001          & 3.977 ± 0.019          \\ \cmidrule(l){2-7} 
                            & $B$ = 4  & 0.893 ± 0.006          & 3.730 ± 0.133          & 3.029 ± 0.010          & 0.948 ± 0.001          & 3.984 ± 0.019          \\
                            & $B$ = 8  & 0.892 ± 0.006          & 3.757 ± 0.128          & 3.031 ± 0.010          & 0.948 ± 0.001          & \textbf{3.987 ± 0.019} \\
                            & $B$ = 16 & \textbf{0.890 ± 0.006} & \textbf{3.758 ± 0.129} & \textbf{3.034 ± 0.010} & \textbf{0.949 ± 0.001} & \textbf{3.987 ± 0.019} \\ \bottomrule
\end{tabular}
\end{table*}

We propose an encoding algorithm for RVQ-based neural codecs, leveraging a beam-search approach to effectively quantize continuous input vectors while mitigating the need for exhaustive exploration of the search space. The main idea is to retain quantization candidates with minimal errors throughout all quantization iterations, selecting the optimal candidate as the final output at the end of the process.
The detailed encoding procedure is outlined in Alg.~\ref{alg:encoding}. 

The encoding process begins with the quantization of the input data $x$ into $B$ codes. This is achieved by identifying the top $B$ closest values to $x$ in the first codebook $C_1$. These $B$ codes are stored in a beam-search candidate list $beam_1$.

Subsequent quantization candidates are determined following the beam-search algorithm. Initially, a residual $r^i_j$ is computed by subtracting the $j$-th candidate of the previous quantization step $beam_{i-1}[j]$ from the input vector $x$. The top $k$ closest values to $r^i_j$ in the $i$-th codebook $C_i$ are then identified and stored in $C_{\text{top}}$. This process iterates $B$ times while generating a pool of $B \times k$ quantization candidates along with their corresponding quantization errors. From this pool, the top $B$ candidates are selected and assigned to the current beam-search candidate list $beam_i$. This iterative refinement occurs for $L-1$ quantization steps ($i$ ranging from $2$ to $L$), progressively enhancing the quantization candidates.

Finally, the best candidate with the minimum quantization error is chosen from $beam_L$. The selected candidate serves as the quantized output $\hat{x}$ produced by our encoding algorithm.

Our algorithm improves the quantization process by incorporating future quantization steps into the determination of the current step's output. It's worth noting that conventional RVQ encoding algorithms represent a special case of our algorithm when $k=1$ and $B=1$, equivalent to a greedy search strategy. In other words, we can obtain quantization outputs with reduced quantization errors by increasing the beam size $B$.

\section{Experiments}
\begin{table}[]
\caption{Comparison of synthesis quality on other audio domains using EnCodec with various beam sizes.}
\label{tab:general}
\centering
\begin{tabular}{@{\hspace{2mm}}clcc@{\hspace{2mm}}}
\toprule
Domain                                                                             &          & Mel Distance ↓          & SI-SNR ↑               \\ \midrule
\multirow{4}{*}{Music}                                                             & $B$ = 1  & 1.060 ± 0.057          & 5.396 ± 0.185          \\ \cmidrule(l){2-4} 
                                                                                   & $B$ = 4  & 1.034 ± 0.057          & 5.688 ± 0.186          \\
                                                                                   & $B$ = 8  & 1.029 ± 0.058          & 5.770 ± 0.187          \\
                                                                                   & $B$ = 16 & \textbf{1.023 ± 0.058} & \textbf{5.835 ± 0.187} \\ \midrule
\multirow{4}{*}{\begin{tabular}[c]{@{}c@{}}Non-Verbal\\ Vocalization\end{tabular}} & $B$ = 1  & 4.339 ± 0.047          & 4.356 ± 0.180          \\ \cmidrule(l){2-4} 
                                                                                   & $B$ = 4  & 4.301 ± 0.047          & 4.773 ± 0.180          \\
                                                                                   & $B$ = 8  & 4.284 ± 0.047          & 4.912 ± 0.180          \\
                                                                                   & $B$ = 16 & \textbf{4.268 ± 0.047} & \textbf{5.014 ± 0.179} \\ \bottomrule
\end{tabular}
\end{table}
\subsection{Experimental setup}
We evaluate the proposed encoding algorithm by employing two neural audio codec models: EnCodec and HiFi-Codec. In all experiments, the official checkpoints are utilized for both models~\footnote{https://github.com/facebookresearch/encodec}\footnote{https://github.com/yangdongchao/AcademiCodec}. The EnCodec checkpoint is pre-trained with a variety of audio sources including  speech, noisy speech, music and general audio. The bit-rate of the EnCodec model is set to 6 kbps in our experiments. Meanwhile, the HiFi-Codec checkpoint is pre-trained with English and Chinese speech data. Both checkpoints are configured for 24 kHz audio.

We conducts the experiments with test data collected from diverse audio sources. Specifically, for speech evaluation, we randomly selected 4000 samples from the test-clean split of LibriTTS~\cite{zen2019libritts}. To assess the algorithm's performance on music, we segment all audio in the test set of MUSDB~\cite{rafii2017musdb18} into 3-second long chunks and randomly choose 3600 samples from them. 
% For general audio evaluation, we sample 200 test data from the test set of AudioSet~\cite{gemmeke2017audio}. 
Additionally, we utilize our internal database for gathering 4000 samples of non-verbal vocalization such as laughs, moans, and sighs.

We rigorously evaluate our proposed algorithm using diverse evaluation metrics. Firstly, we compute the spectral distance between the input data and reconstructed data in mel-spectrogram domains. Our settings include an FFT size of 1024, a hop length of 256, 80 mel-bins, and the utilization of a Hanning window with a size of 1024. The spectral distance is measured in log-scale. For speech data, we assess perceptual quality and intelligibility using PESQ~\cite{rix2001perceptual}, STOI~\cite{taal2010short}, and NISQA~\cite{mittag2021nisqa} scores, along with the scale-invariant signal-to-noise ratio (SI-SNR). The STOI measurement is based on 400 speech samples, ensuring reliable results with a significantly low confidence interval. NISQA is a neural network model specifically trained to estimate the mean opinion score~(MOS) of provided speech data. 
For non-speech domains, we evaluate spectral distance and SI-SNR. 
All experimental results are reported with 95\% confidence intervals.

\subsection{Quantization error}

\begin{table}
\centering
\caption{Comparison of average quantization error.}
\label{tab:qerror}
\begin{tabular}{lcc}
\toprule
         & EnCodec                & HiFi-Codec              \\ \midrule
$B$ = 1  & 5.096 ± 0.018          & 22.29 ± 0.06          \\ \midrule
$B$ = 4  & 4.787 ± 0.017          & 21.96 ± 0.06          \\
$B$ = 8  & 4.693 ± 0.017          & 21.87 ± 0.06          \\
$B$ = 16 & \textbf{4.625 ± 0.017} & \textbf{21.81 ± 0.06} \\ \bottomrule
\end{tabular}
\end{table}

To validate that the quantization output of our algorithm exhibits lower quantization error compared to conventional RVQ encoding, we conducted experiments on the test samples from LibriTTS. We computed the L2 distances between the quantization vectors and the original input signals for each sample, and then averaged these distances to estimate the average quantization error. 

The results are presented in Table~\ref{tab:qerror}. Notably, we observed a consistent decrease in quantization error as the beam size increased, underscoring the superior quantization performance of our algorithm compared to conventional RVQ encoding. We continue to further validate that the reduction in quantization error indeed leads to improved codec performance in the following experiments.

\subsection{Evaluation on speech domain}
% To evaluate the performance of codecs on speech data, we conducted experiments by encoding and decoding 200 test samples from LibriTTS using both EnCodec and HiFi-Codec. 
We compared the conventional RVQ encoding with our proposed algorithm by varying the beam size $B$ over values of 4, 8, and 16. Also, we set the search parameter $k$ to the same value as $B$. The experimental results are detailed in Table~\ref{tab:speech}. 

Both EnCodec and HiFi-Codec demonstrated performance enhancements by integrating the proposed algorithm across all evaluation metrics: mel-spectrogram distance, SI-SNR, PESQ, STOI, and NISQA score. Furthermore, we observed a clear trend that the overall performance improves as the beam size increases across all the evaluation metrics. These findings underscore the efficacy of our proposed algorithm and highlight its superiority over conventional RVQ encoding methods.
% The evaluation scores of both models consistently improved with the increment of the beam size across all the evaluation metrics. 
% HiFi-Codec also exhibited a similar tendency, albeit with the exception of NISQA score. 
% We conjecture that this is because HiFi-Codec already achieves competitive codec performance using conventional RVQ encoding within the speech domain. Nevertheless, across all metrics, the adoption of our algorithm led to improvements over conventional RVQ encoding, supporting the efficacy of our proposed approach.

\subsection{Evaluation on other audio domain}

We extend our evaluation beyond the speech domain to encompass diverse audio categories including music and non-verbal vocalization. In this section, we solely utilized EnCodec for evaluation purposes, as the pre-trained checkpoint of HiFi-Codec is tailored specifically for speech data and may not generalize well to other audio types. 
% The evaluation procedure involved encoding and decoding 200 test samples from MUSDB~(music), and our internal database~(non-verbal vocalization), respectively.

Table~\ref{tab:general} provides a summary of the experimental results across diverse domains and various encoding strategies. Through the integration of our proposed algorithm, we observed enhancements in the mel-spectrogram distance and SI-SNR. It is noteworthy that our algorithm has led to enhanced codec performance even in such challenging scenarios. In addition, the overall performance of our algorithm improves as the beam size $B$ increases, which confirms the effectiveness of our algorithm. These trends align closely with those identified in the speech domain evaluation, indicating consistent improvements across various audio categories.

\subsection{Evaluation with various bit-rates}

\begin{table}[t]
\centering
\caption{Comparison of speech quality using EnCodec with various bit-rates. $N_q$ denotes the number of codebooks used for encoding.}
\label{tab:bit-rate}
{\footnotesize
\begin{tabular}{@{\hspace{2mm}}clccc@{\hspace{2mm}}}
\toprule
Bandwidth                             &          & PESQ ↑                 & NISQA ↑                \\ \midrule
\multirow{2}{*}{3 kbps ($N_q$ = 4)}   & $B$ = 1  & 2.053 ± 0.008          & 3.008 ± 0.022          \\
                                      & $B$ = 16 & \textbf{2.139 ± 0.008} & \textbf{3.098 ± 0.022} \\ \midrule
\multirow{2}{*}{6 kbps ($N_q$ = 8)}    & $B$ = 1  & 2.726 ± 0.010          & 3.491 ± 0.022          \\
                                      & $B$ = 16 & \textbf{2.850 ± 0.010} & \textbf{3.588 ± 0.022} \\ \midrule
\multirow{2}{*}{12 kbps ($N_q$ = 16)} & $B$ = 1  & 3.338 ± 0.011          & 3.811 ± 0.022          \\
                                      & $B$ = 16 & \textbf{3.439 ± 0.011} & \textbf{3.872 ± 0.022} \\ \midrule
\multirow{2}{*}{24 kbps ($N_q$ = 32)} & $B$ = 1  & 3.670 ± 0.011          & 3.992 ± 0.021          \\
                                      & $B$ = 16 & \textbf{3.691 ± 0.011} & \textbf{4.001 ± 0.021} \\ \bottomrule
\end{tabular}
}
\end{table}

% Previous experiments uses the bit-rate of 6 kbps for EnCodec. In this section, we analyze the algorithm efficacy with the various bit-rates of EnCodec: 1.5, 3, 6, 12, and 24 kbps.

In this section, we conduct the analysis of our proposed algorithm's efficacy by exploring its performance across a range of bit-rates for EnCodec. While previous experiments focused on a fixed bit-rate of 6 kbps, we now investigate how our algorithm performs at bit-rates 3, 6, 12, and 24 kbps when applied to EnCodec. We assessed PESQ and NISQA scores using the test samples from LibriTTS, and the results are presented in Table~\ref{tab:bit-rate}. 

Our algorithm consistently exhibits remarkable performance improvements across all configured bit-rates. 
Notably, these enhancements persist even at the higher bit-rate of 24 kbps, where EnCodec uses 32 codebooks. 
This indicates that the sequence of codes derived from conventional RVQ encoding still retains meaningful quantization errors, despite the utilization of a many number of codebooks. 
Our proposed algorithm effectively mitigates these errors and offers a promising solution applicable to RVQ-based neural codec models across a diverse range of bit-rates.

\subsection{Utilization of GPU on proposed algorithm}
% \begin{figure}
%     \centering
%     \hspace*{-0.1in}
%     \includegraphics[width=0.95\linewidth]{inference-speed.pdf}
%     \caption{Comparison of Inference Time on CPU and GPU.}
%     \label{fig:speed}
% \end{figure}

\begin{table}[]
\centering
\caption{Comparison of inference time~(ms) on CPU and GPU.}
\label{tab:speed}
\begin{tabular}{@{\hspace{2mm}}lcc@{\hspace{2mm}}}
\toprule
         & CPU          & GPU                    \\ \midrule
$B$ = 1  & 43.53 ± 0.20 & 6.780 ± 0.008          \\ \midrule
$B$ = 4  & 50.88 ± 0.21 & \textbf{7.248 ± 0.008} \\
$B$ = 8  & 74.03 ± 0.30 & \textbf{7.241 ± 0.008} \\
$B$ = 16 & 167.6 ± 0.51 & \textbf{7.366 ± 0.009} \\ \bottomrule
\end{tabular}
\end{table}

As the beam size increases, the computational cost of our algorithm scales accordingly, potentially leading to decreased efficiency in terms of inference speed. Nonetheless, we can alleviate this issue by employing a parallelizable implementation, particularly focusing on optimizing lines 7 to 16 in Algorithm~\ref{alg:encoding}, and harnessing the computational power of GPUs. This approach ensures that the inference speed remains largely unaffected despite the increase in beam size. To validate this strategy, we conducted experiments encoding 5-second audio segments using an Intel Xeon Gold 6426Y CPU and an NVIDIA RTX 4090 GPU, running each configuration 10,000 times on the GPU and 1,000 times on the CPU.

The inference times are summarized in Table~\ref{tab:speed}. A naive implementation of the proposed algorithm, which does not leverage parallelism, results in a substantial increase in inference time as the beam size grows. For example, as the beam size $B$ varies from $4$ to $16$, the processing time extended by 229\%. In contrast, when we utilize parallel implementation with GPU, the processing time increases by only 1.6\%. Additionally, incorporating our algorithm leads to a modest increase of 6.9\% in inference time, as observed when comparing $B=1$ and $B=4$ in the GPU setting. These results indicate that our algorithm remains efficient even with increasing beam sizes, due to the optimized use of GPU computational power.

\section{Conclusion}
In this work, we introduced the encoding algorithm designed for RVQ-based neural codecs. To illustrate our rationale, we highlighted a key limitation of conventioanl RVQ encoding: the quantization output often exhibits a suboptimal quantization error when compared to the attainable optimal code sequence. To reduce the quantization error, we propose a method wherein quantization candidates with minimal errors are retained across all quantization steps and the best matching candidate is selected as the quantization output at the end of the process. This strategy enables us to obtain a quantization output with reduced error while circumventing the huge computational overhead of exploring all possible code combinations. With EnCodec and HiFi-Codec, we validated the efficacy of our algorithm in diverse audio domains using metrics such as spectral distance, PESQ, STOI, NISQA score, and SI-SNR. We believe that our algorithm can be generally employed to RVQ-based models to further enhance codec performance.

% \pagebreak
\clearpage

\bibliographystyle{IEEEbib}
\bibliography{refs}

\end{document}